\documentclass[
reprint,
superscriptaddress,
amsmath,amssymb,
aps,
]{revtex4-1}

\usepackage{graphicx}
\usepackage{dcolumn}
\usepackage{bm}

\usepackage[utf8]{inputenc}
\usepackage[T1]{fontenc}
\usepackage{mathptmx}

\usepackage{siunitx}
\usepackage{braket}
\usepackage{natbib}
\usepackage[utf8]{inputenc}
\usepackage{graphicx,import} 
\usepackage{verbatim}
\usepackage{dcolumn}
\usepackage{bm}
\usepackage[colorlinks]{hyperref}
\usepackage{xcolor}

\DeclareSIUnit\dBm{dBm}
\DeclareSIUnit{\samplespersecond}{SPS}
\DeclareSIUnit{\bit}{bit}
\DeclareSIUnit{\byte}{B}

\usepackage{tikz}
\usepackage{graphicx}
\usepackage{pgfplots}
\pgfplotsset{compat=newest}
\usepgfplotslibrary{units,groupplots}

\pgfplotsset{colormap={jet}{[1pt]
		rgb(0pt)=(0.0, 0.0, 0.5);
		rgb(1pt)=(0.0, 0.0, 0.535650623885918);
		rgb(2pt)=(0.0, 0.0, 0.589126559714795);
		rgb(3pt)=(0.0, 0.0, 0.62477718360071299);
		rgb(4pt)=(0.0, 0.0, 0.67825311942958999);
		rgb(5pt)=(0.0, 0.0, 0.71390374331550799);
		rgb(6pt)=(0.0, 0.0, 0.76737967914438499);
		rgb(7pt)=(0.0, 0.0, 0.80303030303030298);
		rgb(8pt)=(0.0, 0.0, 0.85650623885917998);
		rgb(9pt)=(0.0, 0.0, 0.90998217468805698);
		rgb(10pt)=(0.0, 0.0, 0.94563279857397498);
		rgb(11pt)=(0.0, 0.0, 0.99910873440285197);
		rgb(12pt)=(0.0, 0.0, 1.0);
		rgb(13pt)=(0.0, 0.0176470588235293, 1.0);
		rgb(14pt)=(0.0, 0.049019607843137254, 1.0);
		rgb(15pt)=(0.0, 0.096078431372549025, 1.0);
		rgb(16pt)=(0.0, 0.12745098039215685, 1.0);
		rgb(17pt)=(0.0, 0.17450980392156862, 1.0);
		rgb(18pt)=(0.0, 0.22156862745098038, 1.0);
		rgb(19pt)=(0.0, 0.25294117647058822, 1.0);
		rgb(20pt)=(0.0, 0.29999999999999999, 1.0);
		rgb(21pt)=(0.0, 0.33137254901960772, 1.0);
		rgb(22pt)=(0.0, 0.3784313725490196, 1.0);
		rgb(23pt)=(0.0, 0.40980392156862744, 1.0);
		rgb(24pt)=(0.0, 0.45686274509803909, 1.0);
		rgb(25pt)=(0.0, 0.50392156862745097, 1.0);
		rgb(26pt)=(0.0, 0.53529411764705859, 1.0);
		rgb(27pt)=(0.0, 0.58235294117647063, 1.0);
		rgb(28pt)=(0.0, 0.61372549019607847, 1.0);
		rgb(29pt)=(0.0, 0.66078431372548996, 1.0);
		rgb(30pt)=(0.0, 0.69215686274509802, 1.0);
		rgb(31pt)=(0.0, 0.73921568627450984, 1.0);
		rgb(32pt)=(0.0, 0.77058823529411768, 1.0);
		rgb(33pt)=(0.0, 0.81764705882352939, 1.0);
		rgb(34pt)=(0.0, 0.86470588235294121, 0.99620493358633777);
		rgb(35pt)=(0.0, 0.89607843137254906, 0.97090449082858954);
		rgb(36pt)=(0.034788108791903853, 0.94313725490196076, 0.93295382669196714);
		rgb(37pt)=(0.060088551549652112, 0.97450980392156861, 0.9076533839342189);
		rgb(38pt)=(0.098039215686274495, 1.0, 0.8697027197975965);
		rgb(39pt)=(0.12333965844402275, 1.0, 0.84440227703984827);
		rgb(40pt)=(0.16129032258064513, 1.0, 0.80645161290322587);
		rgb(41pt)=(0.18659076533839339, 1.0, 0.78115117014547764);
		rgb(42pt)=(0.22454142947501579, 1.0, 0.74320050600885512);
		rgb(43pt)=(0.26249209361163817, 1.0, 0.70524984187223283);
		rgb(44pt)=(0.2877925363693864, 1.0, 0.67994939911448449);
		rgb(45pt)=(0.3257432005060088, 1.0, 0.6419987349778622);
		rgb(46pt)=(0.35104364326375709, 1.0, 0.61669829222011385);
		rgb(47pt)=(0.38899430740037944, 1.0, 0.57874762808349156);
		rgb(48pt)=(0.4142947501581275, 1.0, 0.55344718532574344);
		rgb(49pt)=(0.45224541429475007, 1.0, 0.51549652118912082);
		rgb(50pt)=(0.49019607843137247, 1.0, 0.47754585705249841);
		rgb(51pt)=(0.5154965211891207, 1.0, 0.45224541429475018);
		rgb(52pt)=(0.55344718532574311, 1.0, 0.41429475015812778);
		rgb(53pt)=(0.57874762808349134, 1.0, 0.38899430740037955);
		rgb(54pt)=(0.61669829222011374, 1.0, 0.35104364326375714);
		rgb(55pt)=(0.64199873497786197, 1.0, 0.32574320050600891);
		rgb(56pt)=(0.67994939911448438, 1.0, 0.28779253636938651);
		rgb(57pt)=(0.70524984187223261, 1.0, 0.26249209361163817);
		rgb(58pt)=(0.74320050600885468, 1.0, 0.22454142947501621);
		rgb(59pt)=(0.78115117014547741, 1.0, 0.18659076533839347);
		rgb(60pt)=(0.80645161290322565, 1.0, 0.16129032258064513);
		rgb(61pt)=(0.84440227703984805, 1.0, 0.12333965844402273);
		rgb(62pt)=(0.86970271979759628, 1.0, 0.098039215686274495);
		rgb(63pt)=(0.90765338393421868, 1.0, 0.060088551549652092);
		rgb(64pt)=(0.93295382669196703, 1.0, 0.03478810879190386);
		rgb(65pt)=(0.97090449082858932, 0.95933188090050858, 0.0);
		rgb(66pt)=(0.99620493358633766, 0.93028322440087174, 0.0);
		rgb(67pt)=(1.0, 0.88671023965141638, 0.0);
		rgb(68pt)=(1.0, 0.84313725490196101, 0.0);
		rgb(69pt)=(1.0, 0.81408859840232406, 0.0);
		rgb(70pt)=(1.0, 0.7705156136528688, 0.0);
		rgb(71pt)=(1.0, 0.74146695715323196, 0.0);
		rgb(72pt)=(1.0, 0.69789397240377649, 0.0);
		rgb(73pt)=(1.0, 0.66884531590413965, 0.0);
		rgb(74pt)=(1.0, 0.62527233115468439, 0.0);
		rgb(75pt)=(1.0, 0.58169934640522891, 0.0);
		rgb(76pt)=(1.0, 0.55265068990559207, 0.0);
		rgb(77pt)=(1.0, 0.50907770515613682, 0.0);
		rgb(78pt)=(1.0, 0.48002904865649987, 0.0);
		rgb(79pt)=(1.0, 0.4364560639070445, 0.0);
		rgb(80pt)=(1.0, 0.40740740740740755, 0.0);
		rgb(81pt)=(1.0, 0.36383442265795229, 0.0);
		rgb(82pt)=(1.0, 0.33478576615831535, 0.0);
		rgb(83pt)=(1.0, 0.29121278140886042, 0.0);
		rgb(84pt)=(1.0, 0.24763979665940472, 0.0);
		rgb(85pt)=(1.0, 0.21859114015976777, 0.0);
		rgb(86pt)=(1.0, 0.1750181554103124, 0.0);
		rgb(87pt)=(1.0, 0.14596949891067557, 0.0);
		rgb(88pt)=(1.0, 0.1023965141612202, 0.0);
		rgb(89pt)=(0.99910873440285231, 0.07334785766158336, 0.0);
		rgb(90pt)=(0.94563279857397531, 0.029774872912127992, 0.0);
		rgb(91pt)=(0.90998217468805731, 0.00072621641249104307, 0.0);
		rgb(92pt)=(0.8565062388591802, 0.0, 0.0);
		rgb(93pt)=(0.80303030303030321, 0.0, 0.0);
		rgb(94pt)=(0.76737967914438521, 0.0, 0.0);
		rgb(95pt)=(0.71390374331550821, 0.0, 0.0);
		rgb(96pt)=(0.6782531194295901, 0.0, 0.0);
		rgb(97pt)=(0.6247771836007131, 0.0, 0.0);
		rgb(98pt)=(0.589126559714795, 0.0, 0.0);
		rgb(99pt)=(0.535650623885918, 0.0, 0.0);
		rgb(100pt)=(0.5, 0.0, 0.0);
}}

\begin{document}
	
	\title{State preparation of a fluxonium qubit \\ with feedback from a custom FPGA-based platform}
	
	\author{Richard~Gebauer} 
	\email[Corresponding author: ]{richard.gebauer@kit.edu}
	\author{Nick~Karcher}
	\affiliation{Institute for Data Processing and Electronics, Karlsruhe Institute of Technology, 76131 Karlsruhe, Germany}
	\author{Daria~Gusenkova}
	\affiliation{Physikalisches Institut, Karlsruhe Institute of Technology, 76131 Karlsruhe, Germany}
	\author{Martin~Spiecker}
	\affiliation{Physikalisches Institut, Karlsruhe Institute of Technology, 76131 Karlsruhe, Germany}
	\author{Lukas~Grünhaupt}
	\affiliation{Physikalisches Institut, Karlsruhe Institute of Technology, 76131 Karlsruhe, Germany}
	\author{Ivan~Takmakov}
	\affiliation{Physikalisches Institut, Karlsruhe Institute of Technology, 76131 Karlsruhe, Germany}
	\affiliation{Institute of Nanotechnology, Karlsruhe Institute of Technology, 76344 Eggenstein-Leopoldshafen, Germany}
	\affiliation{Russian Quantum Center, National University of Science and Technology MISIS, Moscow 119049, Russia}
	\author{Patrick~Winkel}
	\affiliation{Physikalisches Institut, Karlsruhe Institute of Technology, 76131 Karlsruhe, Germany}
	\author{Luca~Planat}
	\affiliation{Université Grenoble Alpes, CNRS, Grenoble INP, Institut Néel, 38000 Grenoble, France}
	\author{Nicolas~Roch}
	\affiliation{Université Grenoble Alpes, CNRS, Grenoble INP, Institut Néel, 38000 Grenoble, France}
	\author{Wolfgang~Wernsdorfer}
	\affiliation{Physikalisches Institut, Karlsruhe Institute of Technology, 76131 Karlsruhe, Germany}
	\affiliation{Institute of Nanotechnology, Karlsruhe Institute of Technology, 76344 Eggenstein-Leopoldshafen, Germany}
	\affiliation{Université Grenoble Alpes, CNRS, Grenoble INP, Institut Néel, 38000 Grenoble, France}
	\author{Alexey~V.~Ustinov}
	\affiliation{Physikalisches Institut, Karlsruhe Institute of Technology, 76131 Karlsruhe, Germany}
	\affiliation{Russian Quantum Center, National University of Science and Technology MISIS, Moscow 119049, Russia}
	\author{Marc~Weber}
	\affiliation{Institute for Data Processing and Electronics, Karlsruhe Institute of Technology, 76131 Karlsruhe, Germany}
	\author{Martin~Weides}
	\affiliation{Physikalisches Institut, Karlsruhe Institute of Technology, 76131 Karlsruhe, Germany}
	\affiliation{James Watt School of Engineering, University of Glasgow, Glasgow G12 8LT, United Kingdom}
	\author{Ioan~M.~Pop}
	\affiliation{Physikalisches Institut, Karlsruhe Institute of Technology, 76131 Karlsruhe, Germany}
	\affiliation{Institute of Nanotechnology, Karlsruhe Institute of Technology, 76344 Eggenstein-Leopoldshafen, Germany}
	\author{Oliver~Sander}
	\affiliation{Institute for Data Processing and Electronics, Karlsruhe Institute of Technology, 76131 Karlsruhe, Germany}
	
	\date{\today} 
	
	\begin{abstract}
		We developed a versatile integrated control and readout instrument for experiments with superconducting quantum bits (qubits), based on a field-programmable gate array (FPGA) platform. Using this platform, we perform measurement-based, closed-loop feedback operations with \SI{428}{\nano\second} platform latency. The feedback capability is instrumental in realizing active reset initialization of the qubit into the ground state in a time much shorter than its energy relaxation time $T_1$. We show experimental results demonstrating reset of a fluxonium qubit with \SI{99.4}{\percent} fidelity, using a readout-and-drive pulse sequence approximately \SI{1.5}{\micro\second} long. Compared to passive ground state initialization through thermalization, with the time constant given by $T_1=$~\SI{80}{\micro\second}, the use of the FPGA-based platform allows us to improve both the fidelity and the time of the qubit initialization by an order of magnitude.
	\end{abstract}
	
	\maketitle
	
	The control of superconducting quantum bits (qubits) \cite{TransmonKoch,FluxoniumDevoret,Devoret1169,QuEngGuideScQubits,MwPhotScQuCirc} relies on the ability to read out their state with high precision and apply custom pulses, conditioned on the result of the readout, on a time scale much shorter than the qubit lifetime $T_1$. For the same signal integration time, the readout fidelity can be increased significantly using Josephson parametric amplifiers (JPAs) \cite{QuantumJumps,JPA,DJJAA_2019,ParAmpQuSigJCirc}. The rapid signal processing and decision taking, on the other hand, requires custom hardware based on field-programmable gate arrays (FPGAs), as recently demonstrated in Refs.~\cite{FeedbackDiCarlo,FeedbackCambagneIbarcq,QECOfek}.
	
	Here we present a versatile FPGA-based control and readout platform for experiments with superconducting qubits, which enables fast (compared to $T_1$) feedback loops to control qubits depending on their state. This type of dynamical and conditional control sequences provides the basis for experiments and algorithms relevant for quantum computation, such as active reset \cite{FeedbackDiCarlo} or quantum error correction \cite{QECOfek}.
	
	To perform such closed loops, we employ measurement-based feedback where the qubit, dispersively coupled to a readout resonator \cite{DispReadout,QuantInducShuntedScCircDevoret2016}, is first projected by an initial quantum non-demolition measurement \cite{QNDMeasurement,QNDSuperconducting,projectiveFluxonium2018} onto its ground or excited state, $\ket{0}$ or $\ket{1}$, respectively. The result of this measurement is evaluated in real-time on our platform, by separating the plane of the in-phase and quadrature (IQ) components of the readout resonator response signal in two semi-planes, one attributing the measurement outcome to state $\ket{0}$, the other to state $\ket{1}$. This is achieved by performing linear discriminant analysis which, in case of symmetric and state-independent noise on the response signal, represents also the optimal Bayesian classifier \cite{BeyererPatternRecognition}. Based on the result, different pulse sequences can follow to further manipulate the qubit.
	
	An important figure of merit is the response time or feedback latency of the platform. In the current state of development we report a platform latency of \SI{428}{\nano\second}. This starts to be advantageous in state preparation protocols for typical superconducting qubits, with energy relaxation times $T_1$ exceeding tens of microseconds. The response time is defined as the delay between the last sample of the readout pulse entering the platform and the first sample of the output pulse, which is conditioned on the acquired information. It is platform-specific and therefore does not include experimental parameters such as the propagation delay through the microwave lines connecting to the cryogenic setup, or the duration of the readout pulse. 
	The response time is solely determined by internal delays of the electronics. 
	The main contributions are given by the conversion from the analog to the digital domain and back, as well as by decimation and interpolation filters used to reduce the sample rate in the digital domain.
	We estimate that latencies as low as \SI{150}{\nano\second} are possible with an optimized platform design.
	
\begin{figure}
		\def\svgwidth{\linewidth}
		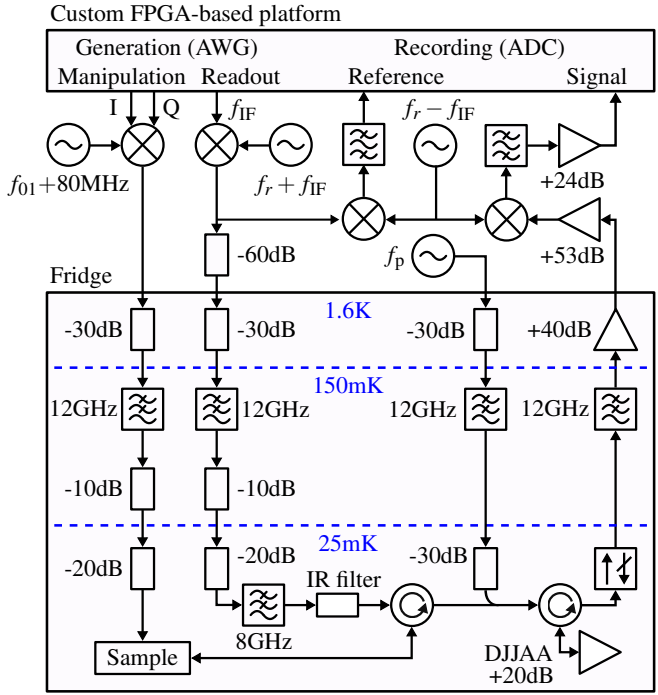
		\caption{\label{fig:setup}Interferometer setup with custom FPGA-based platform for time-domain experiments. The platform offers two IQ channels to generate shaped pulses for the readout and manipulation of the sample. In this setup, only one quadrature of the readout pulse is used. The frequencies of the pulses are $f_\mathrm{IF}=\SI{62.5}{\mega\hertz}$ and $\SI{80}{\mega\hertz}$, respectively. By using a single side-band mixer for the readout and an IQ mixer for the manipulation pulse, the microwave signals are upconverted to the desired resonator and qubit frequencies $f_\mathrm{r}$ and $f_{01}$, respectively. The readout pulse is split before the cryostat input. One branch is directly downconverted and used as reference by the platform. The other branch of the signal is routed through the cryogenic measurement setup, after which it is amplified, downconverted and digitized at the signal input of the platform. From the evaluation of these two signals on the platform we obtain the I and Q components.}
	\end{figure}

	Currently, the platform consists of a Xilinx Zynq UltraScale+ RFSoC ZCU111 Evaluation Board with a XCZU28DR-2FFVG1517E chip. 
	It integrates the FPGA, two processors, eight \SI{6,554}{\giga\samplespersecond} \SI{14}{\bit} digital-to-analog and eight \SI{4,096}{\giga\samplespersecond} \SI{12}{\bit} analog-to-digital converters.
	Our FPGA firmware runs at a clock speed of \SI{125}{\mega\hertz} and handles signals at a sample rate of \SI{500}{\mega\samplespersecond}. 
	Input and output signals are decimated and interpolated from and to \SI{4}{\giga\samplespersecond}, which is the sample rate at which the converters operate. The platform has four output channels, designed to operate as two IQ pairs, one intended for qubit manipulation pulses and one for the readout. The platform further provides two input channels to acquire the readout pulses which passed through the experimental setup. Here, we use them to digitize the signal and reference from an interferometer setup (see experimental schematics in Fig.~\ref{fig:setup}).
	In our case, single side-band mixers are employed and the readout pulse is split after upconversion.
	One branch is directly downconverted and serves as the reference signal.
	The other branch passes through the cryostat and is modulated by the sample before being downconverted as well.
	
	After digitization, the FPGA first compensates the different cable delays by delaying the reference signal, according to an initial calibration.
	It then extracts the I and Q components by interfering the signal with the analytic representation \cite{analyticsignalsmath} of the reference.
	This is achieved by multiplying the signal point-wise with the reference resulting in a value for the I component.
	In a separate multiplication to obtain the Q component, the reference is first shifted by $\pi/2$.
	For intermediate frequencies close to $f_\mathrm{IF} = \SI{62.5}{\mega\hertz}$, this corresponds to a shift of two samples on the FPGA.
	Afterwards the two products are summed up for a recording time of $\SI{800}{\nano\second}$ corresponding to the duration of the readout pulse.
	The obtained values for the I and Q components of the signal are then evaluated in order to infer the state of the qubit.
	Acquisition, IQ calculation, as well as qubit state estimation can run continuously, in parallel and in a pipelined manner.
	By offering the possibility to continue with a different pulse sequence after each qubit state estimation, the platform allows the user to react to this state in real-time. For example, one of the most basic protocols employing this capability is to initialize the state of the qubit on the fly.
	
	Indeed, high-fidelity and rapid state initialization become an essential prerequisite for almost any experimental sequence as the coherence properties of superconducting qubits improve \cite{FluxoniumT1ms,FluxoniumKIT,CoherJQubitMartinis2013,ProtectionScQuEnergyDecay,HighCohJJQu3DcQED,2p5Dtransmon2016,SupprDissipScQuPartPop2014}.
	Generally speaking one can opt to perform either an autonomous stabilization sequence \cite{DrivenResetProtocolDevoret2013, autonomousSuperposition2012}, or to use active reset to initialize the qubit state within a few microseconds, significantly faster than the thermalization time.
	
	To test the feedback capabilities of our platform, we perform active reset on a superconducting granular aluminum fluxonium qubit \cite{FluxoniumKIT} dispersively coupled to a readout cavity. 
	We start by measuring the qubit state with relatively long readout pulses of \SI{800}{\nano\second}, boosted by a dispersion engineered Josephson junction array parametric amplifier (DJJAA) \cite{DJJAA_2019}.
	The state is encoded in the amplitude and phase of the cavity response signal due to the dispersive frequency shift \cite{DispReadout,QuantInducShuntedScCircDevoret2016} of the cavity.
	From this information (which is represented in the IQ-plane) the qubit state is estimated. 
	If the platform detects the excited state $\ket{1}$, a $\pi$ pulse is executed to rotate it to the ground state $\ket{0}$. 
	If state $\ket{0}$ is recognized, no $\pi$ pulse follows. Thereby, after the operation, the qubit is prepared in its ground state.
	
	\begin{figure}
		\begin{tikzpicture}
		\begin{groupplot}[width=0.288\textwidth, height=0.32\textwidth,
		group style={
			group name=myplot,
			group size=2 by 1,
			horizontal sep=0cm,
			vertical sep=0.5em
		},
		legend image code/.code={}, 
		legend pos=south west,
		legend style={font=\normalsize, inner sep=0.2em},
		scaled ticks={real:1e3},
		xtick scale label code/.code={},
		ytick scale label code/.code={},
		grid=major,
		axis on top,
		axis equal,
		xmin=-12261,xmax=15639,
		ymin=-9022,ymax=15904,
		unit markings=parenthesis
		]
		\nextgroupplot[
		ylabel={$Q$}, y unit={arb.~unit},
		every axis x label/.append style={at=(ticklabel cs:1.1)},
		xlabel={$I$}, x unit={arb.~unit},
		]
		\node[anchor=north west] at (axis cs:-10000,19000) {$T_\mathrm{eff} = \SI{30.0}{\milli\kelvin}$};
		\addplot graphics[
		xmin=-10203,xmax=13694,
		ymin=-8347,ymax=15048,
		] {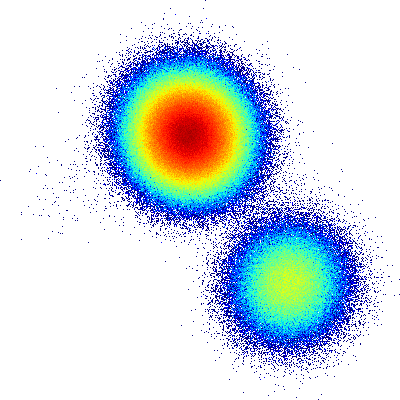};
		\addlegendentry{(a) Equilibrium};
		
		\node[anchor=south west] at (axis cs:5500,11000) {$\ket{0}$};
		\node[anchor=south west] at (axis cs:10500,3000) {$\ket{1}$};
		
		\addplot+[
		black, thick, dashed, no marks, smooth,
		samples=2, domain=-15000:20000
		] {(5923.9689974457215*x + 93309.7735984663)/(8668.53671934866)};
		
		\nextgroupplot[
		point meta min=1,
		point meta max=281,
		yticklabels={,,},
		]
		\node[anchor=north west] at (axis cs:-10000,19000) {$T_\mathrm{eff} = \SI{11.9}{\milli\kelvin}$};
		\addplot graphics[
		xmin=-11131,xmax=14840,
		ymin=-8268,ymax=15904,
		] {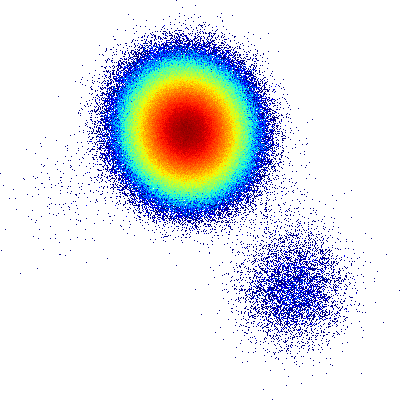};
		\addlegendentry{(b) After reset};
		\end{groupplot}
		
		\begin{axis}[
		at={(myplot c1r1.outer north west)}, anchor=south,
		yshift=1.5cm, xshift=1.25cm,
		hide axis,
		scale only axis,
		height=0pt,
		width=0pt,
		colorbar,
		colormap name={jet},
		colorbar style={
			ylabel={Counts},
			ymode=log,
			log ticks with fixed point,
			height=0.40\textwidth,
			rotate=270,
			ticklabel style={below},
			ytick={1,2,4,10,20,40,100,200},
			ylabel style={
				at={(axis cs:0.5,0.9)},
				anchor=east,
				rotate=-90,
			},
		},
		point meta min=1,
		point meta max=276,
		axis on top,
		rotate=270,
		]
		\end{axis}
		\end{tikzpicture}\vspace{-1cm}
		\caption{\label{fig:areset} Demonstration of quantum feedback by using an active reset sequence. (a) shows the cavity response before the active reset operation while the qubit is in thermal equilibrium with the environment. The linear discriminant used to determine the qubit state in real-time on our platform is indicated as dashed line. (b) shows the response directly after preparing the qubit in its ground state $\ket{0}$ with an active reset operation.}
	\end{figure}
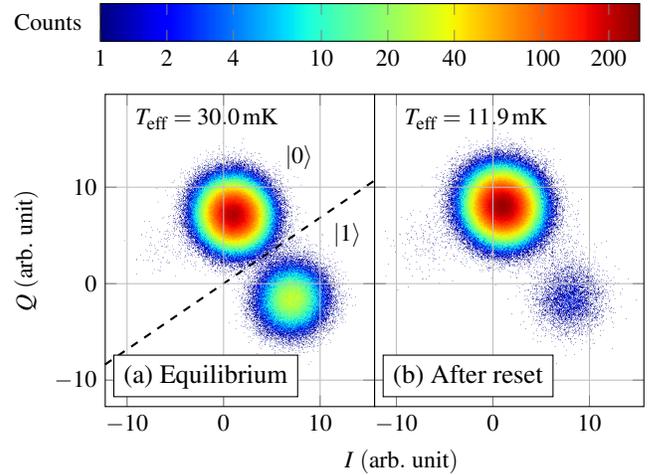
	We report reset fidelities up to \SI{99.4}{\percent}.
	The fidelity is determined by first letting the qubit thermalize and then performing the active reset operation followed by an additional readout to characterize the resulting state.
	The probability to end up in the ground state $\ket{0}$ after the reset is equivalent to the fidelity of the active reset operation.
	Histograms of the response in the IQ-plane of the cavity coupled to the fluxonium qubit, before and after the active reset operation, are shown in Fig.~\ref{fig:areset}.
	Each histogram comprises one million single-shot measurements and was directly obtained by the platform.
	The dispersive frequency shift of the cavity makes the two states distinguishable and allows us to directly extract the qubit state population from the histograms. 
	
	In thermal equilibrium, right before the operation, a $\ket{1}$ state population of \SI{11.7}{\percent} is obtained. Considering the transition frequency $f_{01} = \SI{1.26}{\giga\hertz}$ of the fluxonium qubit at the flux and $T_1$ sweet spot \cite{SupprDissipScQuPartPop2014}, this population corresponds to an effective temperature of \SI{30.0}{\milli\kelvin}, comparable to the temperature of the cryostat mixing chamber. Directly after the reset operation, the $\ket{1}$ state population is only \SI{0.6}{\percent}. We thereby achieved an effective cooling of the qubit to \SI{11.9}{\milli\kelvin}. The achieved reset fidelity is limited by the experimental setup and choice of parameters, setting the state discrimination in the IQ-plane. The Bayesian error alone places an upper bound of \SI{99.5}{\percent} fidelity.
	
	To summarize, our FPGA-based platform enables qubit ground state initialization with \SI{99.4}{\percent} fidelity, 
	using a sequence approximately \SI{1.5}{\micro\second} long, out of which only \SI{428}{\nano\second} account for the intrinsic platform latency. 
	Future efforts will focus on reducing this platform latency down to \SI{150}{\nano\second}, 
	and operating at larger readout powers, which should increase the qubit state separation and decrease the readout pulse duration below \SI{100}{\nano\second} \cite{MuxRoScQubitsHeinsoo2018}.
	
	\begin{acknowledgments}
		Funding was provided by 
		the Initiative and Networking Fund of the Helmholtz Association, within the Helmholtz Future Project ‘Scalable solid state quantum computing’
		and
		the Alexander von Humboldt Foundation in the framework of a Sofja Kovalevskaja award endowed by the German Federal Ministry of Education and Research.
		R. Gebauer acknowledges support by the Helmholtz International Research School for Teratronics (HIRST).
		N. Karcher acknowledges support by the Karlsruhe School of Elementary Particle and Astroparticle Physics (KSETA).
		I. Takmakov and A.V. Ustinov acknowledge support from the Ministry of Education and Science of the Russian Federation in the framework of the Increase Competitiveness Program of the National University of Science and Technology MISIS (contract no. K2-2017-081).
		M. Weides acknowledges support by the European Research Council (ERC) under the Grant Agreement No. 648011, 
		and
		the Deutsche Forschungsgemeinschaft (DFG) projects INST 121384/138-1FUGG.
		Facilities use was supported by the KIT Nanostructure Service Laboratory.
		We acknowledge Qkit \footnote{\url{https://github.com/qkitgroup/qkit}} for providing a convenient measurement software framework.
	\end{acknowledgments}
	
	\nocite{*}
	\bibliography{proceedings-icqt19}
	
\end{document}